# Wi-Fi Gesture Recognition on Existing Devices

**Rajalakshmi Nandakumar, Bryce Kellogg, Shyamnath Gollakota**
University of Washington

**ABSTRACT**

This paper introduces the first wireless gesture recognition system that operates using existing Wi-Fi signals and devices. To achieve this, we first identify limitations of existing wireless gesture recognition approaches that limit their applicability to Wi-Fi. We then introduce algorithms that can classify gestures using information that is readily available on Wi-Fi devices. We demonstrate the feasibility of our design using a prototype implementation on off-the-shelf Wi-Fi devices. Our results show that we can achieve a classification accuracy of 91% while classifying four gestures across six participants, without the need for per-participant training. Finally, we show the feasibility of gesture recognition in non-line-of-sight situations with the participants interacting with a Wi-Fi device placed in a backpack.

**ACM Classification Keywords**
H.5.2. [User Interfaces]: Input Devices and Strategies.

**INTRODUCTION**

Is it possible to leverage Wi-Fi for gesture recognition? Given the ubiquity of Wi-Fi connectivity on mobile devices, a positive answer would allow us to enable gesture interaction on existing devices including laptops, smart TVs, and mobile phones, without additional hardware. More importantly, since these signals do not require line-of-sight and can traverse through material (e.g., cloth), they can enable a number of novel non-line-of-sight gesture interaction applications — e.g., enabling the user to perform in-air gestures at the phone in a pocket or a bag, to say control volume or answer a call.

While researchers have recently made progress in wireless motion detection [3] and gesture recognition [11, 10, 2], prior solutions are limited in that they require custom wireless hardware [11, 3, 10, 2] and cannot operate with existing Wi-Fi signals and devices. In this paper, we introduce *Wi-Fi Gestures*, the first non-light-of-sight gesture recognition solution that can be enabled on existing devices using only a software patch. Our design leverages Wi-Fi packets received on commodity devices to perform gesture recognition in both line-of-sight and non-line-of-sight scenarios.

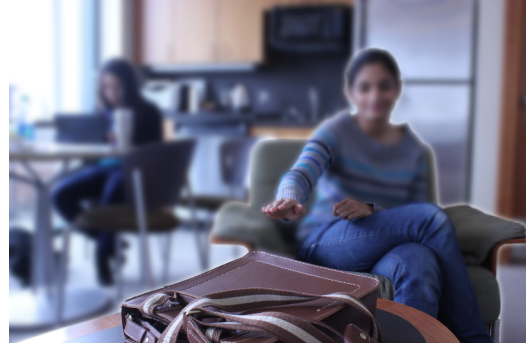

**Figure 1.** Wi-Fi gestures enables wireless gesture recognition using Wi-Fi signals and devices. It can enable interaction with devices in non-line-of-sight scenarios (e.g., when device is in a bag).

The key challenge in achieving the above goal is that traditional techniques used for wireless gesture recognition such as Doppler [11] and Angle-of-Arrival [2] leverage changes in the phase of the wireless signals. Such an approach requires the phase of the wireless carrier generated by the radio hardware to be stable across time. Our experiments show that this is problematic since Wi-Fi devices use low-cost hardware components that do not generate a consistent phase across consecutive packets — hardware with the required phase-consistency is two to three orders of magnitude more expensive. Thus, there is minimal correlation between the phases of successive Wi-Fi packets, rendering phase-based approaches inapplicable for commercial Wi-Fi devices.

To address the above challenge, we introduce a novel algorithm that leverages the Wi-Fi amplitude variations that are readily available on commodity devices in the form of RSSI and CSI information [9]. At a high level, Wi-Fi Gestures detects large amplitude peaks caused by human gestures. Specifically, as the human moves her arm, the wireless reflections from her arm either constructively or destructively interfere with the direct signal from the Wi-Fi transmitter. This results in peaks and troughs in the amplitude of the received signals. Our algorithm uses the size and timing of these peaks to uniquely classify gestures including push, pull, lever, and punch, without the need for per-user training.

We demonstrate the feasibility of our design by building a prototype on off-the-shelf Wi-Fi devices. Our findings about Wi-Fi gestures are as follows:

- It classifies four arm gestures—push, pull, lever, and punch—with an accuracy of 91% across six humans. The accuracy is 89% when the device is in a backpack.

- The above accuracies are independent of the location of the Wi-Fi transmitter and are achieved up to a distance of about one feet between the human and the Wi-Fi receiver.

- The rate of false positive events—gesture detection in the absence of one—is 0.02 events per minute over a 60 minute

period in a busy office environment with thirteen occupants. We achieve this by using a specific repetition start gesture to gain access to the system.

**Contributions.** We introduce the first wireless gesture recognition design that operates on existing Wi-Fi signals and devices. We identify limitations with traditional wireless approaches that limit their applicability to Wi-Fi. We then introduce algorithms that extract gesture information using the amplitude information that is commonly available on Wi-Fi devices. Finally, we build a prototype of our design and demonstrate the feasibility of gesture recognition in non-line-of-sight scenarios, i.e., with the device in a backpack. While the gesture set used in this paper is small, we believe that the algorithmic primitives introduced, e.g., timing and peak variations, could generalize to a broader set of gestures.

## RELATED WORK

Existing gesture-recognition systems can be classified as vision-based, infrared-based, electric-field sensing, ultrasonic, and wearables. The Xbox Kinect, Leap Motion, PointGrab, and CrunchFish use advances in cameras and computer vision to enable gesture recognition. Visible-light base approaches however, by definition, cannot work in non-line-of-sight scenarios. The Samsung Galaxy S4 introduced an "air gesture" feature that uses infrared emitters and detectors for gestures, but is known to be sensitive to lighting conditions [1] and is limited to line-of-sight. Ultrasonic systems such as SoundWave [8] transmit ultrasound waves and analyze them for gesture recognition. However, we are not aware of ultrasonic gesture systems that operate in non-line-of-sight scenarios. Finally, prior work on inertial sensing and other on-body systems require instrumenting the human body with sensing devices [7, 6]. In contrast, we focus on gesture recognition without requiring such instrumentation.

Recent work has leveraged wireless signals for detecting motion such as running [5], and walking forward and backward [3]. WiSee [11], AllSee [10], and WiTrack [2] have also shown the feasibility of extracting gesture information from wireless signals in non-line-of-sight scenarios. These systems however require custom capabilities like ultra-wideband radar transceivers [4, 2], interference-nulling hardware [3], and specialized receiver hardware including USRPs [11] and circuit boards [10]. They have also not been demonstrated to work with Wi-Fi transmissions. In contrast, we are the first to enable gesture recognition using Wi-Fi signals and devices.

## WI-FI GESTURES

Wi-Fi Gestures is a wireless gesture recognition system that operates in non-line-of-sight scenarios and utilizes Wi-Fi packets to perform gesture detection and classification on existing Wi-Fi devices. Achieving this is challenging for two main reasons: First, Wi-Fi is a shared channel, where multiple devices use random access to share the medium. This results in packet transmissions that are not evenly spaced and sometimes even have large gaps in time.

Second, and more important, is the lack of useful phase information. Prior wireless gesture recognition technologies rely on changes in the phase information [11, 2]; the phase information provided by commodity Wi-Fi cards, however, is unstable for use in gesture recognition. To demonstrate this, we used a Dell Inspiron Laptop and an Asus Eee PC (both equipped with Intel 5300 Wi-Fi cards) to transmit and receive Wi-Fi packets, respectively. We measured the phase of the received packets using the Intel CSI Toolkit [9]. Fig. 2 shows the measured phase of each packet over the course of 100 ms. The plot shows that the phase across packets is uncorrelated; this is because unlike software radios (e.g., USRP/WARP) that use expensive oscillators with stable phase, commodity Wi-Fi hardware is orders of magnitude cheaper and hence uses components that have lower stability properties.

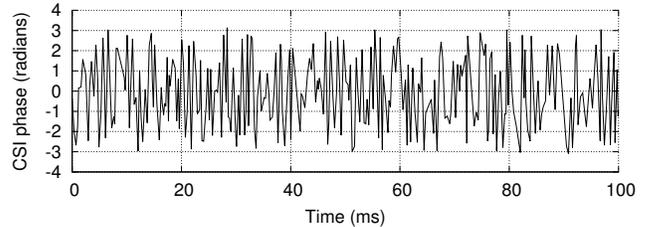

**Figure 2.** Phase of received Wi-Fi packets over the course of 100 ms. The lack of consistency in phase make the use of phase-based approaches difficult for gesture recognition on commodity Wi-Fi devices.

Next, we describe how to perform gesture recognition using only the amplitude of the Wi-Fi channel, i.e., CSI and RSSI. At a high level, this involves three main steps: signal conditioning, peak detection, and gesture classification.

### Signal Conditioning

The goal of signal conditioning is threefold: 1) to account for the uneven arrival of packets caused by the bursty nature of Wi-Fi transmissions, 2) remove any underlying temporal variations in the signal such as glitches or long term changes, and 3) normalize the signal to a common reference.

To account for uneven packet arrival our algorithm fills in small gaps in the Wi-Fi channel samples and interpolates to get evenly spaced samples. To do this, Wi-Fi Gestures uses the 1-D linear interpolation algorithm in MATLAB and obtains 1000 equally spaced samples/s. We note that since the duration of typical human gestures is greater than hundreds of milliseconds, the above interpolation operation preserves the gesture information. We then run the Wi-Fi channel samples through a low-pass filter to reduce noise and glitches. Specifically, since human gestures are relatively slow, we can tune a low pass filter to smooth out the fast varying noise while keeping the slower varying gesture information intact. In our design we use a low-pass filter with the coefficients equal to the reciprocal of one tenth of the number of samples/s.

Finally, minor changes in various environmental factors such as user location, distance between router and device, and objects in the vicinity, can have an impact on the absolute Wi-Fi channel amplitude. The amplitude changes we are interested in for gesture recognition, however, are independent of these absolute values. To extract these changes and remove any bias from the absolute values, we normalize the samples by subtracting a windowed moving average of the channel samples (averaged over 300 ms) from the low-pass filtered ssamples from the previous step. This removes long-term variations and normalizes gesture-free Wi-Fi channel samples to zero.

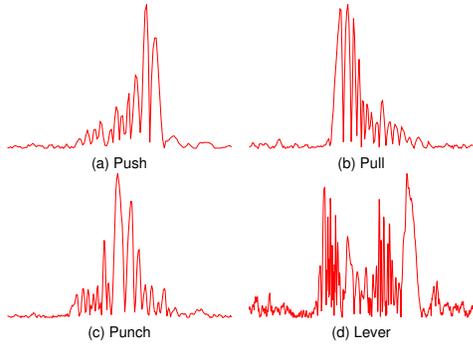

**Figure 3. Each of the four gestures creates a unique pattern of peaks in the conditioned channel samples.**

**Peak Detection Algorithm**
Wi-Fi Gestures detects large peaks and troughs corresponding to constructive and destructive interference cause by human gestures and uses their size and timing to classify gestures. Our peak detection algorithm identifies the size and location of these peaks, but rejects spurious peaks due to noise and glitches. To do this, we use the relative height of a peak to reject peaks that are not part of a gesture. Specifically, any peak above the threshold (1.5 times the mean of the conditioned channel samples) becomes a candidate for inclusion into a gesture. We note that in comparison to ambient human motion such as walking and running, the peaks tend to be higher during intentional gestures close to the device. The above threshold helps reduce confusing them to intentional gestures. To reduce false positives further, we use two main insights: (1) each gesture passes through multiple nodes of constructive/destructive interference, resulting in predictable groups of peaks; and (2) the changes caused by a gesture always result in at least a single large peak that is above the mean noise floor by at least one standard deviation. Using this, we eliminate lone peaks corresponding to glitches and any sets of peaks that do not contain at least one large peak.

**Gesture Classification Algorithm**
Given the sets of peaks corresponding to a gesture, we use their height and timing information to perform classification. To see this, consider a push and a pull gesture. As shown in Fig. 3, the changes cause by the human's arm as she moves it towards the device (a push) result in peaks with increasing heights. In contrast, when the user pulls her hand away from the receiver, the resulting peaks show a decreasing trend in their height. This is because motion closer to the receiver results in larger changes than that farther away. Thus, the wireless changes increase with time as the human moves her hand towards the receiver and they decrease as the human moves her hand away. Similarly, the height of the peaks increases and then decreases for a punch but increase, decrease, and then again increase for a lever gesture. Our classification algorithm hard-codes these patterns to classify the gestures.

**PROTOTYPE IMPLEMENTATION**

**Table 1. Wi-Fi Gestures' Classification Accuracy**

|             | Push  | Pull  | Punch | Lever |
|-------------|-------|-------|-------|-------|
| **On Table**    | 94.03 | 96.97 | 82.76 | 91.89 |
| **In Backpack** | 89.36 | 85.71 | 86.79 | 92.31 |

We implement a software prototype of our design using a Dell Inspiron laptop with an Intel 5300 Wi-Fi card configured to inject Wi-Fi packets at different rates (packets/s) on channel 6 in the 2.4 GHz. The results are similar on all other Wi-Fi channels at 2.4 GHz. To receive these packets we use an Asus Eee PC configured to capture the per-packet CSI data using the CSI Toolkit from [9]. The received CSI values are then processed using the algorithms described earlier.

**EVALUATION AND RESULTS**
We evaluate both the classification accuracy as well as the false positive rate with our prototype.

**Classification Accuracy with Wi-Fi Gestures**
We first evaluate the classification accuracy with six participants (4 males and 2 females) from our organization.

*Experiments:* Participants performed gestures in two scenarios: 1) with the wireless receiver placed on a table directly in front of the participant and 2) with the receiver placed inside a backpack. The participants were asked to sit on a couch in front of the receiver and to perform each gesture in the general direction of the receiver. Before providing gesture data, each participant was shown the desired gestures once and was allowed to practice each gesture three to four times. Each participant then performed each of the four gestures 20 times leaving a three second gap in between gestures, first in the line-of-sight scenario and then again when the receiver was in the backpack. Because each participant was asked to perform a total of 80 gestures in a relatively short span of time, some participants found that their arms got tired while performing the gestures. When we observed that a participant had excessive arm shaking or they mentioned being tired, we asked them to rest their arm until they felt ready to continue. Four of our six participants took at least one rest period during the duration of the experiments.

*Results:* We obtained an average accuracy of 91% and 89% across the six participants when the receiver was in line-of-sight and in the backpack respectively. Table. 1 shows the accuracies for each of the four gestures.

A majority of the misclassifications in our experiments were due to how particular participants performed certain gestures. For example, 40% of the misclassification in the punch gesture in the on-table scenario, were from a single participant who was tired but insisted on not resting. While our current accuracies are promising, we expect them to further improve as participants perform gestures in more relaxed scenarios. We also note that our current algorithm is independent of the participant and uses simple peak variations in the Wi-Fi channel information to perform classification. There are however a number of user-specific features that can improve the classification accuracies. For example, we noticed that three of six participants consistently shaked their arm quite a bit at the end of the push and the pull gestures, resulting in misclassifications. By incorporating this information, one can in principle achieve better accuracies.

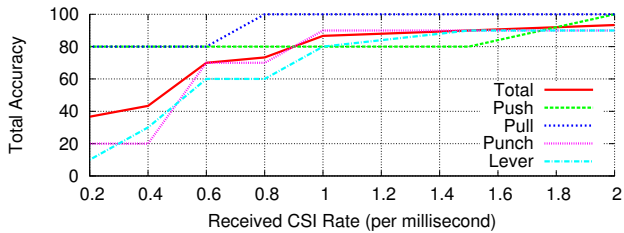

**Figure 4. Effect of CSI Information Rate.** As expected, the accuracies increase with the packet transmission rate.

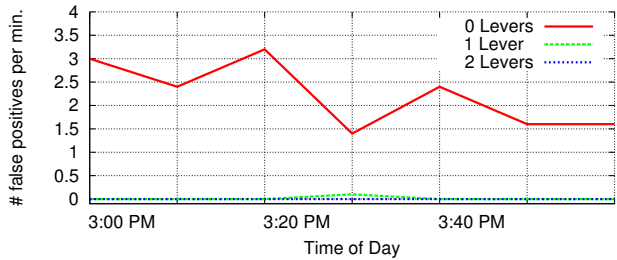

**Figure 5. False Positives from a 60 minute trace.**

Next, we measure the impact of the transmitter position on the accuracies. We pick four randomly chosen transmitter locations including those in a different room as the receiver. Two of our participants agreed to repeat the gestures across these locations. Our results show that the accuracies are not substantially affected by the transmitter location, i.e., they are within 2-3% of each other. This is expected because our algorithm only considers the relative changes in the channel information and not the absolute channel values.

Finally, our accuracies depend on the rate at which the Wi-Fi chipsets provide the channel information. To check this, we set the transmitter to vary the rate at which it transmits packets. We then compute the classification accuracies for our six participants as a function of the rate at which the Wi-Fi receiver provides the CSI samples. Fig. 4 shows that the accuracies increase with the packet transmission rate. We note that while our current design leverages channel information from a single Wi-Fi transmitter, since typical networks have multiple devices, one can in principle combine the CSI information across transmitters to reduce the need for transmitting a large number of packets from a single device.

**False Positives with Wi-Fi Gestures**
Since we leverage the effect of human motion on wireless signals, it is conceivable that random human movement near the receiver may result in spurious gesture detection. To reduce this effect, we implement a start gesture, where the user performs a unique gesture sequence to activate the system before it enters normal detection mode. In our design we use a lever gesture as our start sequence.

*Experiments:* We place our Wi-Fi transmitter and receiver in a busy office space shared by 13 people. The transmitter was placed in a central location in the office and the receiver on a participant's desk. We collect data over a period of 60 minutes while the participants continued to perform normal duties such as type, eating, and moving around.

*Results:* Fig. 5 plots the average number of false gestures as a function of time. The results show that when the receiver does not use a start sequence, the average number of false positive events is about 2.3/min. This is low despite running it next to the participant since our algorithm is designed to only account for large peaks that are characteristic of intentional gestures. The average number of false positives reduces to 0.02 and 0.0/min when a single and double lever gesture is used. This is because a lever motion creates periodic short bursts of amplitude peaks, with a specific range of periodicities, that are unlikely to occur with typically activities.

## CONCLUSION
This paper introduces the first wireless gesture recognition design that operates using existing Wi-Fi signals and devices. Leveraging our design, we demonstrate the feasibility of non-line-of-sight gesture interaction on commodity devices. We believe that the algorithmic primitives introduced in this paper would enable a broader set of gestures than those considered in this paper and also generalize to other widely available signals including cellular transmissions. Given the ubiquity of Wi-Fi on mobile devices, we believe that this paper takes a significant step towards always-available interaction.

**Acknowledgements.** This research is funded in part by UW Commercialization Gap Fund, NSF, and University of Washington.